# *Spatiotemporal cascading of dielectric waveguides.*

*Victor Pacheco-Peña[1], and Nader Engheta[2]*

[1] School of Mathematics, Statistics and Physics, Newcastle University, Newcastle Upon Tyne, NE1 7RU, United Kingdom
[2] Department of Electrical and Systems Engineering, University of Pennsylvania, Philadelphia, PA 19104, USA

email: victor.pacheco-pena@newcastle.ac.uk, engheta@seas.upenn.edu

**Photonic time interfaces, as the temporal analogue of spatial interfaces between two media, consist of a rapid change of the electromagnetic properties of a material (such as permittivity, $\varepsilon$, and permeability, $\mu$) while the wave is present in the material. Here we exploit cascading of such time interfaces in spatially cascaded guided-wave structures such as slab waveguides and ring resonators by considering that the relative permittivity of the cladding of dielectric waveguides is rapidly changed at different moments of time from $\varepsilon_{clad\_1}$ to $\varepsilon_{clad\_2}$, while the material of the core remains unchanged in time. It is shown how such time-dependent cladding can enable frequency conversion within the space-time dielectric ring resonator and slab waveguides due to an induced modification of the effective refractive index of the mode propagating within such photonic device. Cascaded frequency conversion is achieved in such cascaded space-time dielectric waveguides and ring resonators, showing how the combination of space and time interfaces can offer further opportunities for manipulation of light-matter interaction using four-dimensional (4D) photonic structures.**

# Introduction

Understanding how to tailor and control light-matter interactions has been of great interest to the scientific and technological communities. This has enabled the proposal and demonstration of groundbreaking applications ranging from nanoscale sensing and antennas, to new imaging techniques and computing with waves, among others[1–3]. The manipulation of light can be achieved by carefully designing spatial inhomogeneities (with isotropic/anisotropic electromagnetic (EM) responses) placed along its propagation path, which is at the core of the photonics and metamaterials research and innovation[4,5]. Recently, the scientific community has also focused attention into exploring scenarios involving wave propagation within time or space-time modulated media (i.e., a four-dimensional (4D) manipulation of waves). One of the hot topics in this context has been the study of media having EM properties ($\varepsilon$, permittivity, or µ, permeability) that are periodically modulated in time or in space-time, demonstrating how they can be used for exotic physical phenomena such as time-modulated meta-atoms, Fresnel drag, nonreciprocity, wave amplification and thermal emission[6–15]. Additionally, there is an increasing interest to study wave propagation in materials where their properties are rapidly changed in time. Such time modulation was first studied several decades ago in the EM realm[16,17] considering a monochromatic EM wave propagating inside an unbounded medium with relative EM parameters $(\varepsilon_1, \mu_1)$ for times $t < t_1$ (all values larger than 1). Then, at $t = t_1$, the EM parameters were changed to $(\varepsilon_2, \mu_2)$ (again, all larger than 1). This type of time modulation is the temporal analogue of a spatial interface between two materials, i.e., a time interface/boundary. While there are analogies between both spatial and temporal scenarios, the physics of time interfaces are different as wavenumber *k* and wavelength $\lambda$ do not change, while the frequency is modified during this process. Time interfaces have been recently studied and explored in intriguing applications such as time reversal and phase conjugation[18,19], nonreciprocity and loss compensation[20,21], space-time media[22,23], temporal anisotropy[24–27], temporal gain and loss[28,29], time gratings[30], topology via space-time crystals[31,32], time modulation in quantum systems[33,34], short pulsed metamaterials[35,36], antireflection coatings and filters[37–40], frequency conversion[41,42], generalization of the Kramers-Kronig relations[43,44], the implication of negative values of $\varepsilon_2$ or $(\varepsilon_2, \mu_2)$[45,46], as well as the exploration of effective medium concepts in 4D [47–51]. Remarkably, recent efforts in this field have enabled the first experimental demonstrations of time interfaces[18,52,53] and their implementation to scenarios such as frequency conversion[54], broadband temporal coherence[55] and the time analogue of the double-slit diffraction experiment[56], showcasing the different opportunities and possibilities that time and space-time media can open.

While most of these studies consider time-interfaces applied in the whole space where the wave propagates, the combination of space and time interfaces along with effective medium concepts can also be exploited to enable a full spatiotemporal manipulation of light-matter interactions[19,47,57,58]. Motivated by the interesting physical phenomena arising from time and space-time modulated media, here we study wave propagation in time modulated dielectric waveguides and ring resonators where only one of their materials is time-dependent. Specifically, we consider dielectric waveguides having the core and cladding materials defined by $\varepsilon_{core} = constant$ and $\varepsilon_{clad}(t)$, respectively. The time-dependent cladding has a permittivity of $\varepsilon_{clad\_1}$ for time $t < t_1$ and it is rapidly changed to $\varepsilon_{clad\_2}$ at $t = t_1$ (with a fall/rise time smaller than the period $T_1$ of the incident signal to induce a time interface). An in-depth study of the proposed space-time dielectric waveguide is presented showing how frequency conversion can be achieved. This is due to the fact that the effective refractive index of the mode propagating inside such space-time waveguide is modified in time when inducing the time interface at $t = t_1$ via the time-dependent cladding. This concept is then applied to ring resonators (as an analogue of an "unbounded media" where the high-Q structure, while being finite in size, can house the wave for a reasonably long time) demonstrating how carefully engineered cascaded ring resonators filled with time-dependent claddings can enable cascaded frequency conversion. Importantly, this is done by implementing the same time-dependent cladding for all the ring resonators but considering that $\varepsilon_{clad}(t)$ is carefully modified from $\varepsilon_{clad\_1}$ to $\varepsilon_{clad\_2}$ at a different moment in time for each resonator; i.e., cascaded ring resonators both in space and time.

## Results and discussion

**Dielectric waveguide with time-dependent cladding: theory**

The schematic representation of the proposed space-time dielectric slab waveguide is shown in Fig. 1a. Throughout the present work we assume that the input EM signal has a free-space wavelength of $\lambda_1 = 1550$ nm ($f_1 \sim 193.54$ THz), i.e., telecommunication wavelength. The slab waveguide consists of a core material having a thickness $d$ and relative permittivity similar to that of Silicon (Si) at the designed telecom wavelength, $\varepsilon_{core} = (3.47)^2$. The core is placed between two dielectric media working as the cladding. The permittivity of the cladding is $\varepsilon_{clad}(t)$, i.e., a time-dependent function (in all these studies we consider non-magnetic materials with relative permeability $\mu_{core} = \mu_{clad} = 1$). For $t < t_1$, the core and the cladding have relative permittivity values of $\varepsilon_{core\_1} = (3.47)^2$ and $\varepsilon_{clad\_1}$, respectively. At $t = t_1$, a time interface is applied to the cladding by rapidly changing its permittivity to $\varepsilon_{clad\_2}$ ($\varepsilon_{clad\_1} \neq \varepsilon_{clad\_2}$), while the core remains unchanged ($\varepsilon_{core\_1} = \varepsilon_{core\_2} = \varepsilon_{core}$). See Fig. 1c for a schematic representation of $\varepsilon_{clad}(t)$. The theoretical analysis of the proposed space-time dielectric waveguide can be divided into two interrelated parts considering the EM parameters of the materials before and after inducing the time interface. For $t < t_1$, the dispersion relation of the dielectric slab waveguide with EM parameters $\varepsilon_{core}$ and $\varepsilon_{clad\_1}$ can be written as the following well-known transcendental equation[4]:

$$\tan\left(\frac{d}{2}\sqrt{\omega_1^2 \varepsilon_{core} - \beta^2}\right) = \frac{\sqrt{\beta^2 - \omega_1^2 \varepsilon_{clad\_1}}}{\sqrt{\omega_1^2 \varepsilon_{core} - \beta^2}} \tag{1}$$

with $\omega_1 = 2\pi f_1$ as the angular frequency at $f_1$ (frequency of the incident signal) and $\beta$ as the effective propagation constant of the whole dielectric waveguide. The value of $\beta$ can be used to calculate an effective permittivity/refractive index of the waveguide for $t < t_1$ as $\varepsilon_{eff\_1} = n_{eff\_1}^2 = \left(\frac{\beta c}{\omega_1}\right)^2$ with $c$ as the velocity of light in vacuum. Now, once the time interface is rapidly induced at $t = t_1$ via a change of $\varepsilon_{clad}(t)$ to $\varepsilon_{clad\_2}$, the dispersion relation from Eq. 1 is transformed to the following expression:

$$\tan\left(\frac{d}{2}\sqrt{\omega_2^2 \varepsilon_{core} - \beta^2}\right) = \frac{\sqrt{\beta^2 - \omega_2^2 \varepsilon_{clad\_2}}}{\sqrt{\omega_2^2 \varepsilon_{core} - \beta^2}} \tag{2}$$

with $\omega_2 = 2\pi f_2$ as the angular frequency at the new frequency $f_2$ induced by the time interface. By comparing Eq. 1-2 one can notice that, as the time interface is only applied to the cladding, it is the effective propagation constant $\beta$ of the mode propagating within the waveguide that must be preserved through the process, similar to[19,47]. Hence, for $t > t_1$ the frequency of the wave traveling inside the space-time dielectric waveguide, $f_2$, can be calculated by forcing $\beta$ to be the same in both Eqs. 1-2. From this new $f_2$, as in the case discussed in Eq. 1, an effective permittivity/refractive index (i.e., effective mode index) for $t > t_1$ can be simply calculated as $\varepsilon_{eff\_2} = n_{eff\_2}^2 = \left(\frac{\beta c}{\omega_2}\right)^2$. Finally, it is important to remark that in the proposed scenario from Fig. 1a, the new frequency $f_2$ ($t > t_1$) will be related to the initial frequency $f_1$ ($t < t_1$) as follows:

$$n_{eff\_1} f_1 = n_{eff\_2} f_2 \tag{3}$$

which is the same expression for conventional time interfaces applied to an unbounded medium [with isotropic, homogeneous changes of $\varepsilon(t)$ ][16,17] but now considering effective values of permittivity/refractive index (i.e., effective mode index). To evaluate the frequency conversion in the structure shown in Fig. 1a, the analytical results (calculated using Eq. 1-2) of the ratios $f_2/f_1$ and $n_{eff\_1}/n_{eff\_2}$ as a function of $\varepsilon_{clad\_1}$ and $\varepsilon_{clad\_2}$ are shown in Fig. 1d,e, respectively. The thickness of the core is $d = 160$ nm, which is chosen to ensure the coupling of only the fundamental TE$_0$ mode at the designed frequency $f_1 \sim 193.54$ THz[4]. By comparing Fig. 1d,e, one can notice how they are the same for all values of $\varepsilon_{clad\_1}$ and $\varepsilon_{clad\_2}$, corroborating Eq. 3 (the values of $\varepsilon_{eff\_2}$ are shown in the Supplementary Materials for completeness). The black dashed line in Fig. 1d,e represents the case when no time interface is applied, meaning that $\frac{f_2}{f_1} = \frac{n_{eff_1}}{n_{eff_2}} = 1$. To further study these results, two examples, $\varepsilon_{clad\_1} = 4$ and $\varepsilon_{clad\_1} = 2$, were extracted from Fig. 1d,e and the results for $f_2/f_1$ and $n_{eff\_1}/n_{eff\_2}$ as a function of $\varepsilon_{clad\_2}$ are shown in Fig. 1f. Here, the star symbols represent the values of $f_2/f_1$ calculated using numerical simulations via the time domain solver of COMSOL Multiphysics®. These numerical results were obtained by exciting the structure with a narrow-band Gaussian signal modulated at the frequency $f_1$ and recording the $E_y$-field distribution at the output of a space-time dielectric slab waveguide of length $l \sim 9\lambda_1$ for $t > t_1$. Then the frequency spectrum was calculated from the recorded $E_y$-field distribution using build-in Fourier transform from COMSOL Multiphysics®. (See further details about the numerical setup in the methods section below and the full numerical solutions in the Supplementary Materials). As observed from Fig. 1f, there is an excellent agreement between the theoretical and numerical calculations, demonstrating how the ratios between frequencies and effective refractive index are smaller for when $\varepsilon_{clad\_1} = 2$, as expected. Importantly, note that as the signal is considered at the output of the waveguide, it is the forward (time refracted) signal that is recorded. Animations showing the case when $\varepsilon_{clad\_1} = 2$ or $\varepsilon_{clad\_1} = 4$ and $\varepsilon_{clad\_2} = 6$ are presented as Supplementary Materials. In those results, a backward wave (time reflected) is also excited as expected, but it is almost negligible in amplitude due to the small change of effective relative permittivity of the waveguide as shown in Fig. 1f. Interestingly, by observing these animations, one also notices some radiated signals being excited once the time interface is applied. This is an expected result that can be explained by considering Eqs. 1-2. For $t < t_1$ the fundamental mode is excited and propagates within the waveguide. Then, once the time interface is applied to the cladding at $t = t_1$, higher-frequency modes (with the same $\beta$) will also be exited. These modes will have larger values of $\omega_2$, for which the wavenumber in the cladding will be larger than $\beta$, and therefore will leak out of the structure and radiate away, similar to[19]. As an example, following Eqs. 1-2, when a time interface is applied such that $\varepsilon_{clad\_1} = 4$ and $\varepsilon_{clad\_2} = 1$, the frequency of the fundamental mode and first higher-frequency mode are $f_2 \sim 203$ THz and $f_2 \sim 360$ THz, respectively. A similar performance has been also reported in[57,58] where it has been shown how a space-time interface can generate radiated fields at the spatial interfaces between time-dependent and time-independent regions. From now on, we focus our attention into the fundamental mode TE$_0$.

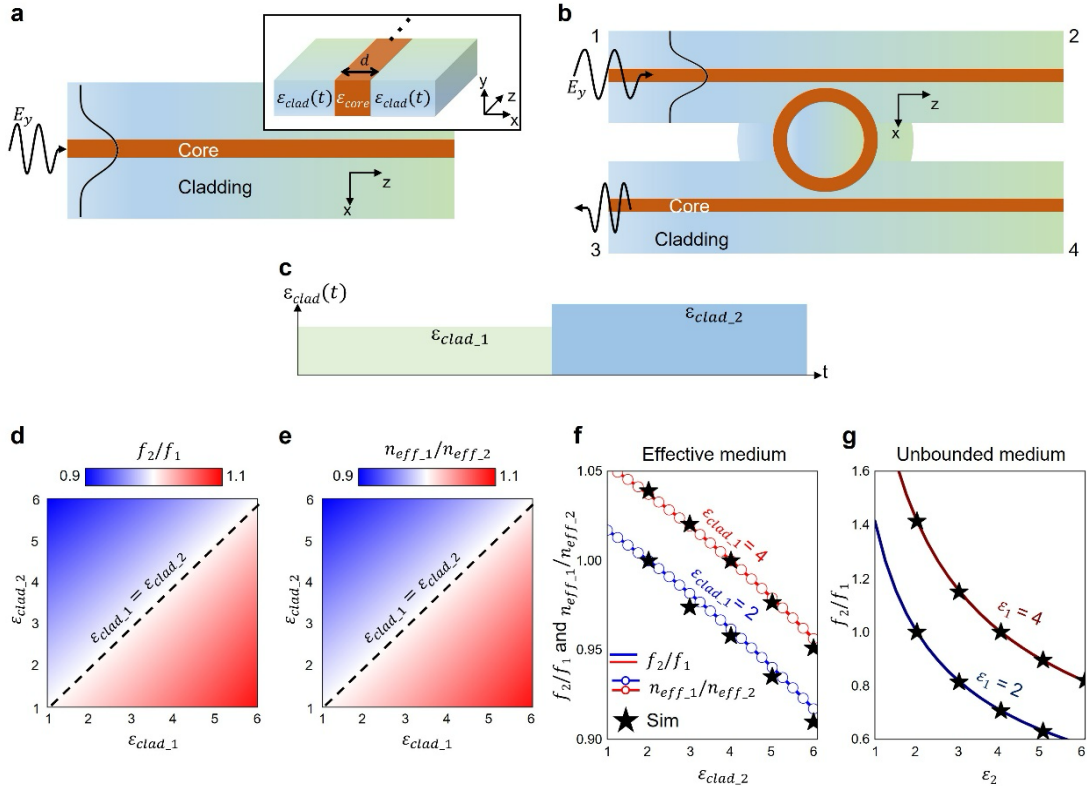

**Fig. 1| Schematic representation of the proposed space-time dielectric waveguide. a,b** space-time waveguide and ring resonator consisting of a core with a time-independent relative permittivity $\varepsilon_{core}$ and a time-dependent cladding with relative permittivity $\varepsilon_{clad}(t)$. For $t < t_1$, $\varepsilon_{clad}(t) = \varepsilon_{clad\_1}$. At $t = t_1$, it is rapidly changed to $\varepsilon_{clad\_2}$. **c,** schematic representation of $\varepsilon_{clad}(t)$. **d,e** analytical results of the ratios $f_2/f_1$ and $n_{eff\_1}/n_{eff\_2}$, respectively, calculated using Eqs. 1-3 as a function of $\varepsilon_{clad\_1}$ and $\varepsilon_{clad\_2}$. $\varepsilon_{core} = (3.47)^2$ in all the calculations. **f,** analytical results of the ratios $f_2/f_1$ (solid lines) and $n_{eff\_1}/n_{eff\_2}$ (white circles) extracted from **d,e** for $\varepsilon_{clad\_1} = 4$ (red plots) and $\varepsilon_{clad\_1} = 2$ (blue plots). **g,** $f_2/f_1$ considering a conventional time interface applied to the entire unbounded medium where the wave is traveling, as a function of $\varepsilon_2$ (relative permittivity for $t > t_1$) considering two values of $\varepsilon_1$ (relative permittivity for $t < t_1$), $\varepsilon_1 = 4$ (dark red line) and $\varepsilon_1 = 2$ (dark blue line). The numerical simulation results are shown as star symbols in panels **f,g**.

For completeness, and to fully understand the implications of the proposed space-time dielectric waveguide in terms of frequency conversion, the analytical and numerical results of the ratios $f_2/f_1$ and $n_1/n_2$ considering a conventional time interface applied to an unbounded medium where a monochromatic wave is traveling [EM parameters changed from $\varepsilon_1$ ($n_1$) to $\varepsilon_2$ ($n_2$) at $t = t_1$] are shown in Fig. 1g. By comparing Fig. 1f and Fig. 1g, one can observe how the ratios, and hence the new frequency of the wave for $t > t_1$, are smaller for the space-time dielectric slab waveguide as compared to the conventional unbounded medium. This is an expected result because only the cladding and not the entire medium is time-dependent in the results shown in Fig. 1f (i.e., an effective medium approach). However, as it will be discussed below, one can exploit cascaded space-time waveguides and/or ring resonators (as shown in Fig. 1b) to further increase the converted frequency. We are particularly interested in ring resonators in the context of space-time interface, because in high-Q ring resonators the monochromatic wave is trapped for a relatively long time (i.e., many cycles depending on the Q of the system), and therefore the rapid temporal change of permittivity of claddings may not have to have the stringent condition of being a fraction of signal period, potentially enabling the implementation of slow temporal modulations as studied in[59].

**Cascading space-time dielectric waveguides**

As discussed in Fig. 1, a single space-time dielectric slab waveguide can be exploited to achieve frequency conversion as the time interface applied to the cladding can indeed create a change of the effective refractive index (i.e., mode index) of the waveguide. What would happen if several of these waveguides were cascaded? Here, we discuss how space-time waveguides can be cascaded (as shown in Fig. 2a) and evaluate their performance for frequency conversion. As shown in Fig. 2a, the first waveguide (waveguide 1, WG1) is used as input. This waveguide has a core width of $d_1 = 160$ nm (as the waveguide used in Fig. 1). The materials for the core and the cladding are the same for all the three cascaded waveguides, meaning that $\varepsilon_{core}$ is time-independent and $\varepsilon_{clad}(t)$ is time-dependent. To account for a space-time cascaded configuration of waveguides, in addition to being cascaded in space (as shown in Fig. 2a), the dimension *time* should also be considered. To achieve this, one can introduce cascaded time interfaces by changing the permittivity of the cladding for each waveguide from $\varepsilon_{clad\_1}$ to $\varepsilon_{clad\_2}$ at different times (*cascading in time*). The time at which the induced time interface should be applied for each waveguide is chosen such that the signal is traveling inside of that waveguide. In this context, the materials for the cladding in all the waveguides can be considered the same, but it is the moment of time at which they are time-modulated that will now be different. In the case shown in Fig. 2, initially, $\varepsilon_{clad\_1} = 4$ for all the waveguides. This value is then rapidly changed to $\varepsilon_{clad\_2} = 1.5$ at $t = t_1, t = t_2,$ and $t = t_3,$ for WG1, WG2 and WG3, respectively [see specific time-dependent functions of $\varepsilon_{clad}(t)$ for each WG in the Supplementary Materials and values in the caption of Fig. 2]. Now, there is a key engineering aspect that one should also consider: once the first time interface is applied to the cladding of WG1 at $t = t_1$, the narrowband modulated pulse inside this waveguide will change its frequency from $f_1$ to $f_2$ following Eqs. 1-3. This pulse will then encounter WG2, i.e., a spatial interface. In order to avoid spatial reflections at this spatial interface, one should carefully engineer the dimension of the core for WG2 ($d_2$) to enable transverse impedance matching between the two dielectric waveguides. As we are working with non-magnetic media (where the permeability is the same in all the materials, and hence it is the same in all the dielectric waveguides) transverse impedance matching for the fundamental mode can be achieved by ensuring that the effective refractive index (i.e., mode index) of the fundamental mode for WG2 should match the effective refractive index of WG1 at the frequency of the new incident pulse $f_2$. To do this, once the new frequency $f_2$ and effective refractive index $n_{eff\_2}$ is calculated for $t_1 < t < t_2$ using Eqs. 1-3, these parameters can be fed back into Eq. 1 to calculate the thickness $d_2$ of the core for WG2. In so doing a spatial reflection will be minimized. It is important to note that changing the dimension $d_2$ of the core for WG2 will then introduce a change of cross-sectional shapes between WG1 and WG2. However, provided that the two core dimensions $d_1$ and $d_2$ are not too different, reflections will still be minimized, meaning that the consideration of equal effective mode index between WG1 and WG2 is a good approximation for impedance matching. Similarly, when the second time interface is applied at $t = t_2$ to the cladding of WG2, the pulse travelling within this waveguide will change its frequency to $f_3$, meaning that to reduce spatial reflections when this pulse reaches WG3, the thickness of the core $d_3$ will also need to be carefully calculated following the same process as the one described for $d_2$. For completeness, the numerical results of the transmitted and reflected spectra for the spatial interfaces between WG1-WG2 and WG2-WG3 are presented in the supplementary materials, demonstrating negligible spatial reflections. The resulting cascaded waveguide geometry is shown in Fig. 2a.

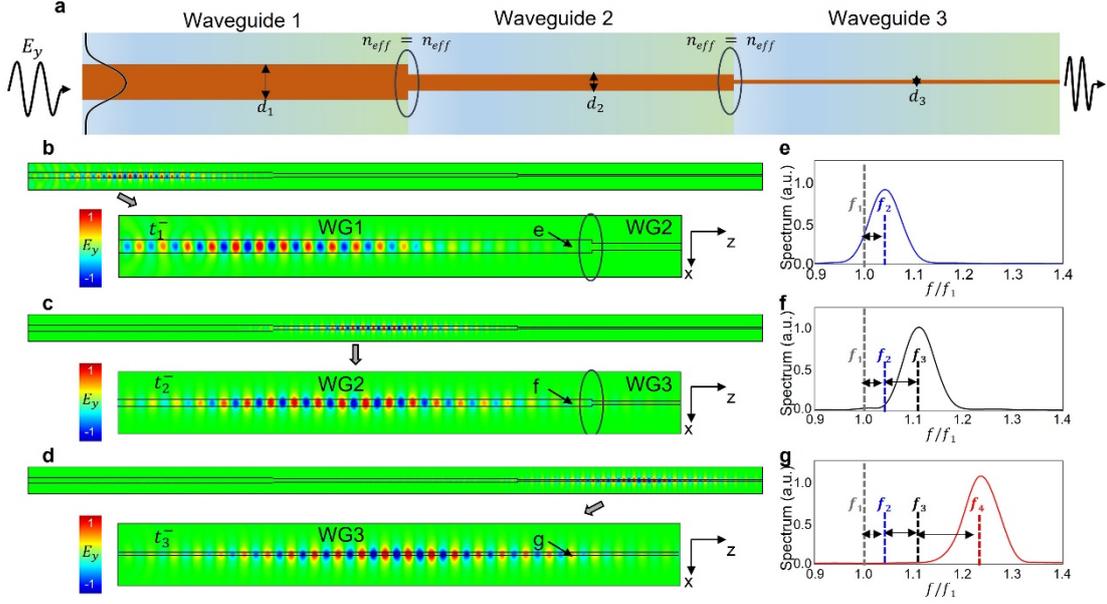

**Fig. 2| Spatiotemporal cascading dielectric waveguides. a,** schematic representation of three cascaded waveguides showing the input and output $E_y$-field distribution. The waveguides (WG1-3) have a core thickness of $d_1 = 160$ nm, $d_2 \sim 127$ nm and $d_3 \sim 93$ nm. The relative permittivity of the core is the same for all the waveguides with a value of $\varepsilon_{core}$ as the results shown in Fig. 1. The cladding of all the waveguides is time-dependent, with an initial and final value of $\varepsilon_{clad\_1} = 4$ and $\varepsilon_{clad\_2} = 1.5$. However, the time interface is applied to each of the claddings at different time instances. For WG1-3, it is changed to $\varepsilon_{clad\_2}$ at $t = t_1 \sim 30T_1$, $t = t_2 \sim 56.1T_1$ and $t = t_3 \sim 81.2T_1$, respectively, with $T_1$ as the period of the incident signal for times $t < t_1$. **b-d,** numerical results of snapshots of the $E_y$-field distribution at time instants before each of the time interfaces is introduced showing the pulse traveling within each of the waveguides. The top panels from **b-d** show the results for the full three waveguide structure while the bottom panels are zoom in from the top panels for WG1, WG2 and WG3, respectively. **e,g** Spectral response of the recorded numerical results for $E_y$-field distribution calculated at the output of each waveguide.

With this configuration, the numerical results for the *y*-polarized (out-of-plane axis) $E_y$-field distribution within WG1-3 at times $t = t_1^-, t = t_2^-$ and $t = t_3^-$, respectively, (i.e., a time instant before each of the time interfaces are applied) are shown in Fig. 2b-d, respectively. As observed, the pulse is inside each of the waveguides when each of the time interfaces is applied. Moreover, no noticeable spatial reflections are observed, as expected due to the carefully designed dimensions of each of the cores. Finally, the simulation result for the $E_y$-field distribution was recorded at the output of each waveguide (as shown by the arrows in Fig. 2b-d) and the spectra are presented in Fig. 2e-g for each waveguide, respectively. From Fig. 2e, after the cladding is changed in time inside WG1 at $t = t_1$, the central frequency of the signal is changed to $f_2/f_1 \sim 1.048$ (analytical value from Eqs. 1-3 of $f_2/f_1 = 1.045$). When the time interface is induced to the cladding of WG2 (Fig. 2f), the central frequency is modified to $f_3/f_1 \sim 1.12$ (or equivalent to $f_3/f_2 \sim 1.068$, in agreement with the analytical value of $f_3/f_2 \sim 1.063$ from Eqs. 1-3). Finally, when the cladding of WG3 is changed in time, the frequency is changed to $f_4/f_1 \sim 1.23$ (i.e., $f_4/f_3 \sim 1.098$, again in-line with the analytical value of $f_4/f_3 \sim 1.096$ from Eqs. 1-3). These results demonstrate how cascaded space-time dielectric slab waveguides could have the potential for cascaded frequency conversion.

**Cascading space-time photonic ring resonators**

Photonic ring resonators are key in many photonics applications such as in circuits and high-speed computing with light[60]. Wavelength conversion in ring resonators has been demonstrated in[41] by modulating the EM properties of the cavity. Here, we consider that only the cladding is time-dependent as in Fig. 1-2. However, effective refractive index approach can still apply here (as discussed in Eqs. 1-3) as the waveguides forming the ring have different materials for the core and cladding, as expected, and the time interface will be applied to the cladding only. The schematic representation of the proposed configuration is shown in Fig. 3a. It consists of three cascaded ring resonators. Each of them formed of a ring of radius $R$ (namely $R_1$, $R_2$ and $R_3$, respectively) and two waveguides: one working as the input (top waveguides) and one as drop (bottom waveguides). The relative permittivity of the core is time-independent with the same value in Fig. 1-2, while the cladding of each resonator is time-dependent, rapidly changing from $\varepsilon_{clad\_1} = 4$ to $\varepsilon_{clad\_2} = 1.5$ at $t = t_1, t = t_2,$ and $t = t_3$, for resonators 1-3, respectively. See schematic of the time-dependent $\varepsilon_{clad}(t)$ for each resonator in Fig. 3b. The core of all the drop waveguides and rings have a width of $d_1 = 160$ nm (as in Fig. 1-2), this is to reduce the number of parameters to be modified and also to enable a smaller change of width of the input waveguides compared to the values discussed in Fig. 2.

Now, the dimensions of the input waveguides and the radius of the rings are carefully engineered following a similar process as the one discussed for Fig. 2. As the incident signal with a central frequency of $f_1$ is applied from the input waveguide of resonator 1, the radius of its ring, $R_1$, is selected to enable maximum coupling of this $f_1$ to the first resonator (see the frequency domain scattering parameters of this first resonator in the Supplementary Material). Then, at $t = t_1$, the cladding of ring resonator 1 is changed in time, inducing a time interface and modifying the frequency of the signal to $f_2$. This signal will then travel towards resonator 2. Similarly to the design from Fig. 2a, the dimension $d_2$ of the input waveguide (top waveguide) for resonator 2 can be calculated using Eqs. 1-3 to achieve refractive index matching, minimizing spatial reflections. The next step is then to select the radius of the ring for resonator 2, $R_2$, to enable the coupling of the incoming signal with a frequency $f_2$ (see Supplementary Materials for the frequency domain scattering parameters as a function of $R_2$). Once the signal is traveling inside the ring resonator 2, the relative permittivity of the cladding for resonator 2 is rapidly changed at $t = t_2$, transforming the frequency of the signal to a new frequency, $f_3$. Next, the same process to calculate the dimensions for resonator 3 can be applied. The new frequency $f_3$, along with the condition of refractive index matching, is implemented to calculate the dimension $d_3$ of the input waveguide of resonator 3 while $R_3$ is engineered to enable the coupling of this signal into the resonator (see Supplementary Materials). However, different from the results shown in Fig. 2, here one should notice that the width of the core of ring 2 and its drop waveguide is $d_1$ rather than $d_2$, meaning that Eq. 1 should be implemented using the parameters $d_1$ and $f_3$. Finally, once the signal with frequency $f_3$ is traveling within resonator 3, the permittivity of the cladding is changed in time at $t = t_3$, changing the frequency of the signal to its final value of $f_4$.

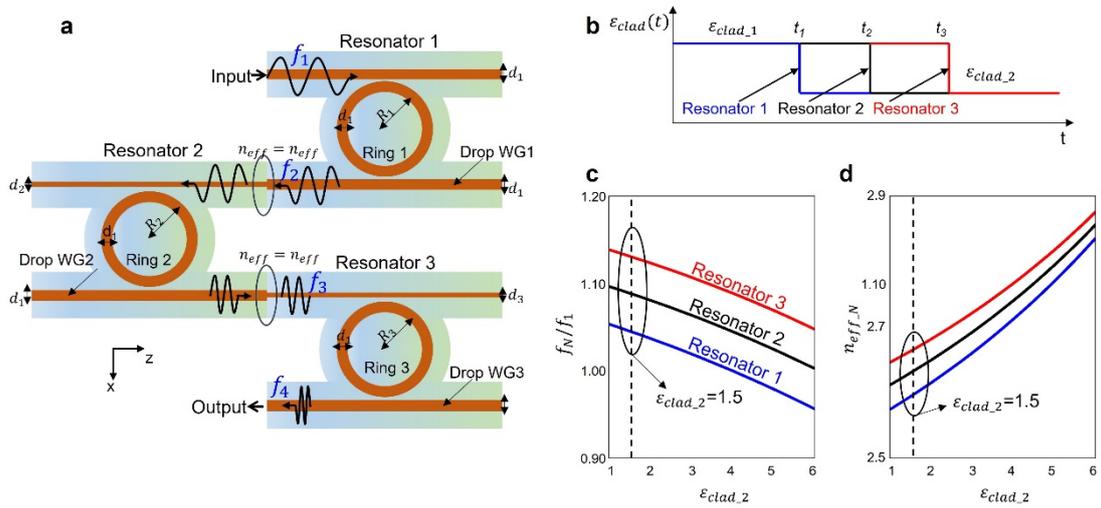

**Fig. 3| Spatiotemporal cascading ring resonators. a,** schematic representation of three cascaded ring resonators. The radius of the rings for each resonator is $R_1, R_2$ and $R_3$, respectively. The width of the core for all the rings and all the drop waveguides (bottom waveguides) is $d_1 = 160$ nm while the input waveguides (top waveguides) have core with $d_1, d_2$ and $d_3$ for resonator 1, 2, and 3, respectively. The relative permittivity of the core is the same for all the waveguides $\varepsilon_{core}$ as in Fig. 1-2. The cladding of all the waveguides is time-dependent, with an initial and final value of $\varepsilon_{clad\_1} = 4$ and $\varepsilon_{clad\_2} = 1.5$. The time-dependent $\varepsilon_{clad}(t)$ for each resonator is schematically shown in **b. c,d,** analytical results of the ratio $f_N/f_1$ (with subscript $N$ =2,3,4 representing each of the frequencies as schematically shown in **a**). **d,** effective refractive index $n_{eff\_N}$. The results from **c,d** have been obtained using Eqs. 1-3 with the final dimensions as shown in Fig. 4.

With this configuration, the numerical results of a design considering three cascaded space-time dielectric ring resonators are shown in Fig. 4. The dimensions and specific time interfaces are detailed in the caption of the figure, showing how the width of the input waveguides are larger than those from the cascaded space-time waveguides discussed in Fig. 2. As explained above, this is a result of forcing the core of the rings and drop waveguides to be the same, $d_1$. This can potentially make their future experimental fabrication more viable and could also allow more cascaded space-time rings than just using cascaded space-time waveguides. Snapshots of the $E_y$-field distribution at different times are shown in Fig. 4a-c considering time instances before each of the time interfaces is applied, respectively. The time-dependent functions for the cladding for each resonator are also plotted in Fig. 4d-f, respectively, showing how the values are changed from $\varepsilon_{clad\_1} = 4$ to $\varepsilon_{clad\_2} = 1.5$ in all the cases, but the change is carried out at different times in order to induce cascaded time interfaces. The $E_y$-field distribution is then recorded at the output of each drop waveguide for each resonator and the spectral response is shown in Fig. 4g-i. As observed, once the first time interface is applied to resonator 1 at $t = t_1$, the frequency is changed to $f_2/f_1 \sim 1.042$ (analytical value of $f_2/f_1 = 1.045$). After the time interface at $t = t_2$ induced to the cladding of resonator 2, the frequency is changed to $f_3/f_1 \sim 1.09$ (equivalent to $f_3/f_2 \sim 1.046$ in line with the analytically calculated value from Eqs. 1-3 of $f_3/f_2 = 1.042$). Finally, after the time interface at $t = t_3$ in resonator 3, the final frequency is $f_4/f_1 \sim 1.13$ (or $f_4/f_3 \sim 1.037$, also in agreement with the analytically calculated value $f_4/f_3 \sim 1.039$). For completeness, the analytical results of the ratio of frequencies $f_N/f_1$ (with $N$ =2,3,4 referring to each of the frequencies from Fig. 3a and Fig. 4) along with the effective refractive index $n_{eff\_N}$ are shown in Fig. 3c,d, respectively. An animation showing the results presented in Fig. 4 can also be found as a Supplementary Material. It is worth noticing how the final frequency after all the time interfaces is slightly smaller for the space-time cascaded rings compared to the space-time waveguides discussed in Fig. 2. This is an expected result due to the fact that the width of the cores of ring resonators and the drop waveguides in Fig. 3-4 have been forced to be $d_1$ in order to avoid smaller values of $d_2$ and $d_3$, compared to those obtained in the structure from Fig. 2. These results demonstrate how cascaded space-time dielectric waveguides could have the potential for cascaded frequency conversion, opening further opportunities for photonic applications using time interfaces. As mentioned before, the scientific community has recently started to report the first experimental validations of time interfaces including water waves, microwaves up to the optical regime. We hope that our proposed space-time waveguides and ring resonators concepts for cascaded frequency conversion will be considered for experimental demonstration in the near future by leveraging the knowledge and techniques of these recently reported experimental work and those efforts currently under development by the scientific community worldwide.

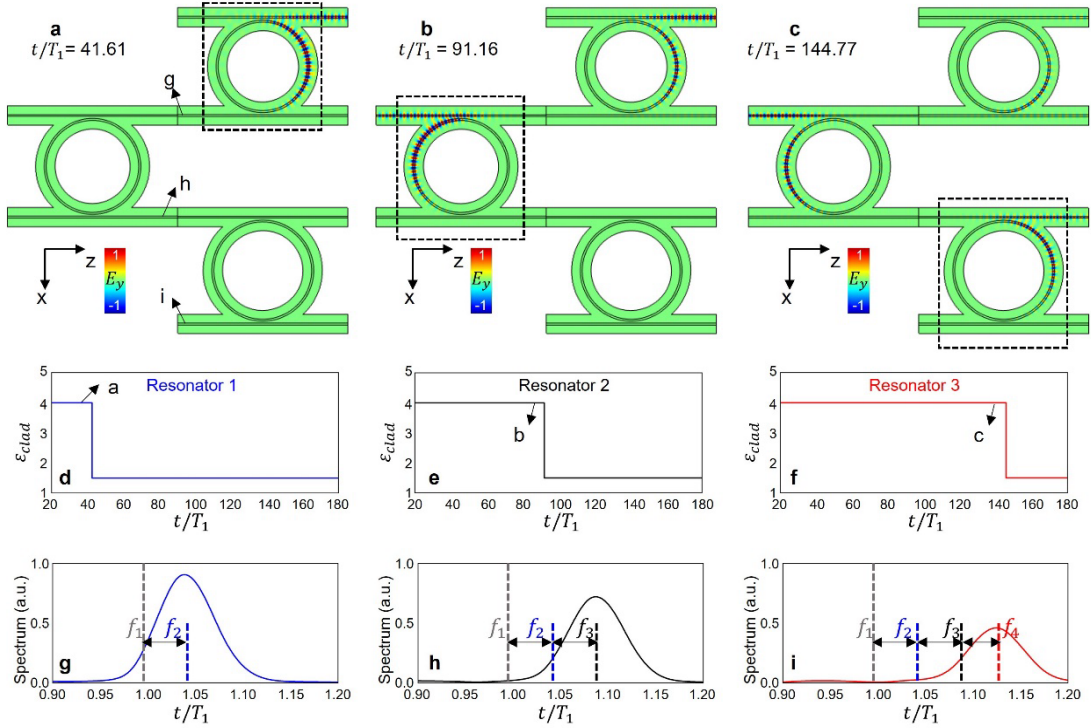

**Fig. 4| Cascading space-time ring resonators: results.** Following the process described in Fig. 3a, the final structure has dimensions $R_1 \sim 3.75$ μm, $R_2 \sim 3.89$ μm and $R_3 \sim 4.1$ μm with core widths of $d_1 = 160$ nm, $d_2 \sim 127$ nm, and $d_3 \sim 129$ nm. **a-c,** snapshots of the numerical result for the $E_y$-field distribution at times before the time interfaces within the resonators is applied. The time-dependent cladding for each resonator is shown in **d-f**, respectively where $\varepsilon_{clad}$ is changed to $\varepsilon_{clad\_2}$ at $t = t_1 \sim 42.5T_1$, $t = t_2 \sim 92T_1$ and $t = t_3 \sim 145.6T_1$, respectively, with $T_1$ as the period of the incident signal for times $t < t_1$. **g-i,** spectrum of the numerical results of the $E_y$ at the output of each drop waveguide for each resonator, demonstrating cascaded frequency conversion.

## Conclusions

In conclusion, we have proposed and discussed the performance of photonic time interfaces applied to all dielectric waveguides and ring resonators. This has been achieved by considering that their core is time-independent but their cladding is changed in time. A full theoretical and numerical study has been carried out demonstrating how frequency conversion is achieved due to a change in the effective refractive index (i.e., effective mode index) of the space-time waveguide when the permittivity of the cladding is rapidly modified in time from $\varepsilon_{clad\_1}$ to $\varepsilon_{clad\_2}$. Cascaded frequency conversion in cascaded space-time dielectric waveguides and ring resonators has also been discussed. This has been achieved by selecting the dimensions of the cascaded waveguides/ring resonators and carefully selecting the time instances at which the cladding of each waveguide/ring resonator is modified (i.e., temporal cascading). All dielectric waveguides and cavities such as ring resonators are key for photonic systems at optical frequencies. Hence, we envision that the work presented here may open further possibilities and opportunities for photonic applications in both space and time, enabling higher degrees of freedom to control light-matter interactions with photonic devices.

**Methods**

All the numerical simulations in Fig. 1-4 were carried out using the time-domain solver of the commercial software COMSOL Multiphysics®. The space-time waveguides used in this work were considered as a rectangular geometry of length $l \sim 9\lambda_1$ and width $d_1 = 160$ nm (as the width of the core as shown in Fig. 1-4) immersed within another rectangle of the same length but width of $10d_1$. In this way, the core and cladding could be assigned their corresponding values of permittivity. Scattering boundary conditions were applied in all directions of the simulation domain with the left boundary being used to excite the structures. For the simulations shown in Fig. 1f, Fig. 2-4, a narrowband incident modulated Gaussian signal was implemented having a standard deviation of $6T_1$ and a central frequency of $f_1 \sim 193.54$ THz). For the results shown in Fig. 1g, i.e., time interfaces applied within an unbounded medium, a similar setup as the one implemented in[37] was used, with dimensions scaled to $\lambda_1$ of the present work. The time interfaces were defined using built-in step functions with a transition time of $\sim 0.015T_1$ and two continuous derivatives. Finally, a minimum mesh size of $\sim 1.2 \times 10^{-6} \lambda_1$ was used to enable accurate results. For the designs shown in the Supplementary Materials section 5, the frequency domain solver of COMSOL Multiphysics® was implemented using a parameter sweep optimization of the parameters $R_1, R_2$ and $R_3$ considering monochromatic incident signals at the frequencies $f_1, f_2$ and $f_3$, respectively from the results shown in Fig. 3-4.

**Acknowledgements**

V.P.-P. would like to thank the support of the Leverhulme Trust under the Leverhulme Trust Research Project Grant scheme (RPG-2020-316). N. E. acknowledges partial support from the Simons Foundation/Collaboration on Symmetry-Driven Extreme Wave Phenomena (grant 733684). For the purpose of Open Access, V. P.-P. has applied a CC BY public copyright license to any Author Accepted Manuscript (AAM) version arising from this submission.

## Conflicts of interests

The authors declare no conflicts of interests.

## Data availability

The datasets generated and analyzed during the current study are available from the corresponding author upon reasonable request.

# Supplementary Materials

## *Spatiotemporal cascading of dielectric waveguides.*

*Victor Pacheco-Peña[1], and Nader Engheta[2]*

[1] *School of Mathematics, Statistics and Physics, Newcastle University, Newcastle Upon Tyne, NE1 7RU, United Kingdom*
[2] *Department of Electrical and Systems Engineering, University of Pennsylvania, Philadelphia, PA 19104, USA*

email: victor.pacheco-pena@newcastle.ac.uk, engheta@seas.upenn.edu

1. $\varepsilon_{eff\_2}$ for a space-time dielectric waveguide: theoretical results

2. Numerical results for space-time dielectric waveguide: single waveguide

3. Time-dependent $\varepsilon_{clad}$ for spatiotemporally cascaded waveguides.

4. Reflection and transmission for spatiotemporally cascaded waveguides.

5. Spatiotemporally cascaded ring resonators: *R* dimensions

6. Supplementary videos:

    a. Supplementary Movie 1: examples of space-time dielectric waveguides.

    b. Supplementary Movie 2: animation of the cascaded space-time ring resonator from Fig. 3-4 of the main text.

# 1. $\varepsilon_{eff\_2}$ for a space-time dielectric waveguide: theoretical results

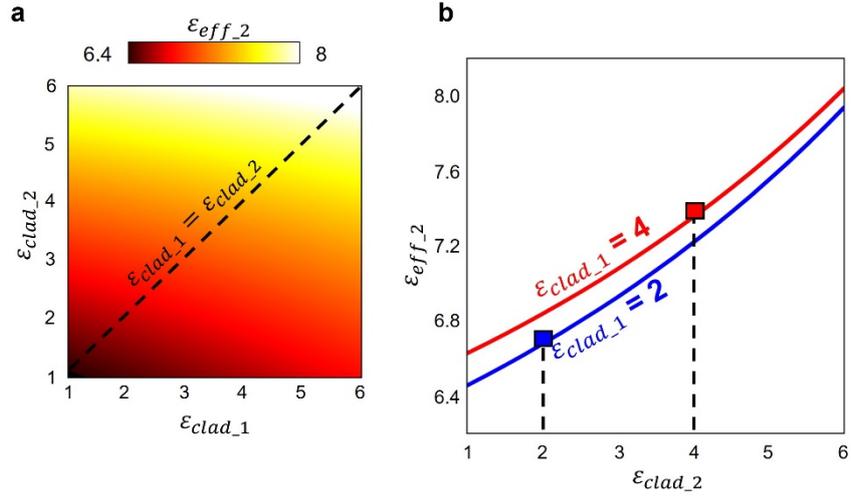

**Fig. S1|** Theoretical calculations of $\varepsilon_{eff\_2}$ considering the space-time dielectric waveguide discussed in Fig. 1a,d,e of the main text. **a,** $\varepsilon_{eff\_2}$ as a function of $\varepsilon_{clad\_1}$ and $\varepsilon_{clad\_2}$. **b,** $\varepsilon_{eff\_2}$ as a function of $\varepsilon_{clad\_2}$ for $\varepsilon_{clad\_1} = 4$ (red line) and $\varepsilon_{clad\_1} = 2$ (blue line) extracted from **a**. The square symbols represent the cases when $\varepsilon_{clad\_1} = \varepsilon_{clad\_2}$, meaning that $\varepsilon_{eff\_1} = \varepsilon_{eff\_2}$.

## 2. Numerical results for space-time dielectric waveguide: single waveguide

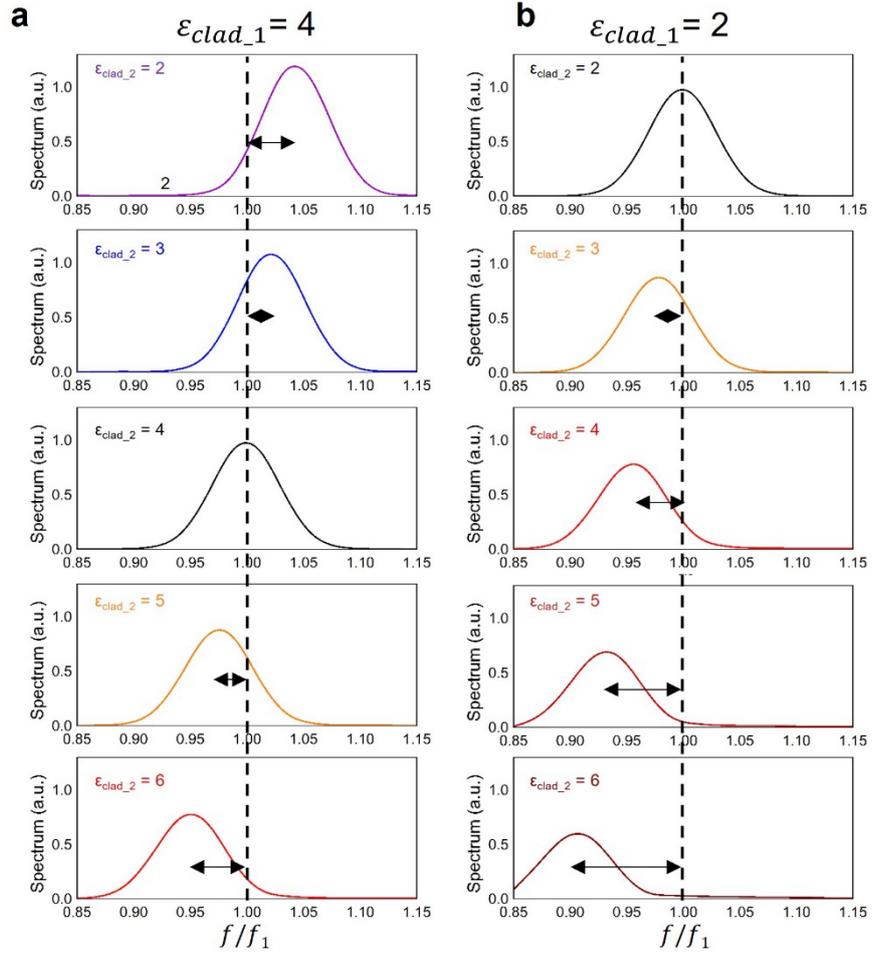

**Fig. S2|** Numerical results of the output spectra for times $t > t_1$ (after the permittivity of the cladding is changed in time) for a single space-time dielectric waveguide excited by a narrowband Gaussian pulse modulated at the central wavelength $\lambda_1 = 1550$ nm. The permittivity of the cladding is **a**, $\varepsilon_{clad\_1} = 4$ or **b**, $\varepsilon_{clad\_1} = 2$ for $t < t_1$. At $t = t_1$ it is changed to different values of $\varepsilon_{clad\_2}$ as shown in the different rows from **a,b**. The central frequency of the spectra from these results correspond to the star symbols shown in Fig. 1f of the main text.

## 3. Time-dependent $\varepsilon_{clad}$ for spatiotemporally cascaded waveguides.

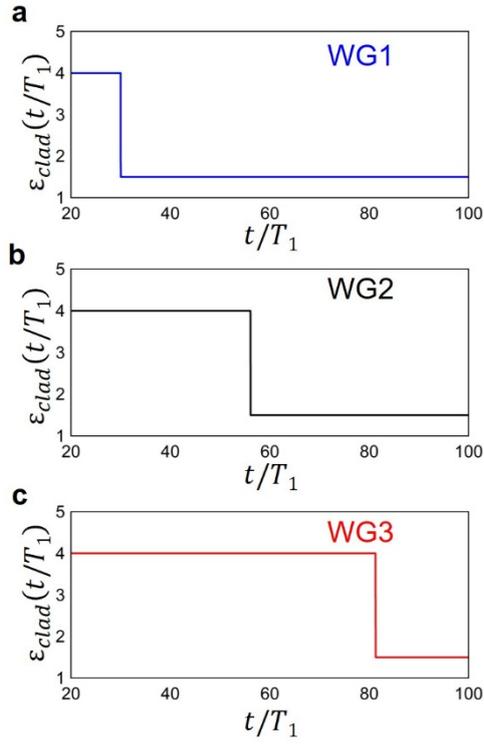

**Fig. S3|** Time-dependent functions of $\varepsilon_{clad}$ for the spatiotemporally cascaded dielectric waveguides discussed in Fig. 2 of the main text. WG1-3 represent waveguides 1-3, respectively.

## 4. Reflection and transmission for spatiotemporally cascaded waveguides.

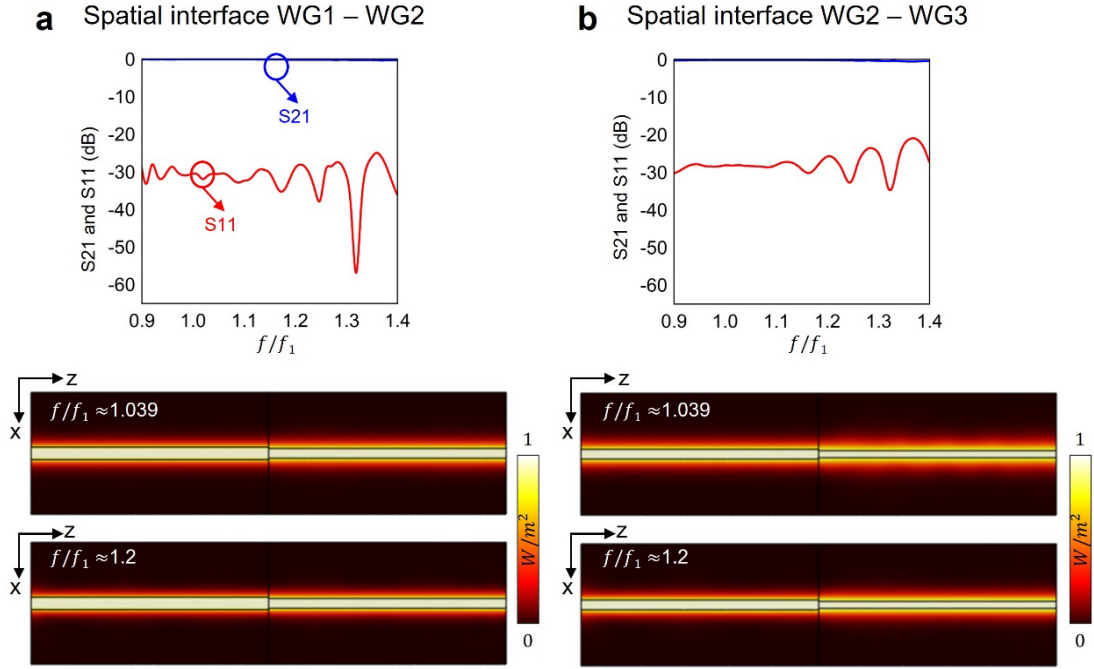

**Fig. S4|** Transmitted and reflected signals at the interfaces between WG1 and WG2 (a) and between WG2 and WG3 (b) for the spatiotemporally cascaded waveguides from Fig. 2 of the main text. The power distribution at two different frequencies are shown at the bottom panels. These results demonstrate that spatial reflections are minimized with the designed structures.

## 5. Spatiotemporally cascaded ring resonators: *R* dimensions

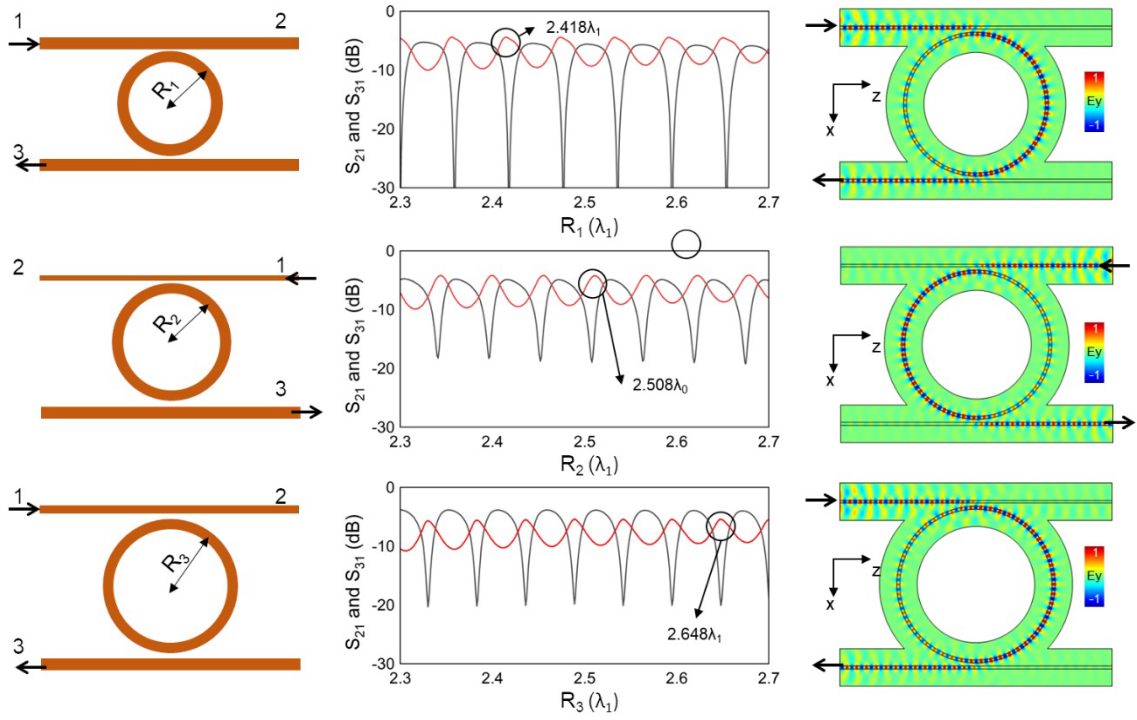

**Fig. S5|** Selection of the dimensions $R_1$, $R_2$ and $R_3$ for the structure shown in Fig. 3 of the main text: (First column) schematic representation of each ring resonator. The final dimension of each ring is marked as a black circle in the second column. (Third column) The simulation results for the $E_y$-field distribution for each case using the dimension chosen from the second column.